\documentclass[dvips,12pt]{iopart}
\usepackage{graphicx}

\begin{document}

\title[Performing a robust Ryan test]{Performing a theoretically robust Ryan test using intermediate-mass black holes}

\author{J S Graber}

\address{LOC, 407 Seward Square SE, Washington, DC 20003}
\ead{ jgra@loc.gov,jgraber@mailaps.org}
\begin{abstract}
We investigate the possibility of using binary IMBH inspirals to perform the Ryan test of general relativity in a theoretically robust manner using data from early in the detectable part of the inspiral.  We find this to be feasible and compute the masses of the most favourable systems. 
\end{abstract}


\submitto{Class. Quantum Grav.}

\section{Introduction: Using LISA to challenge Einstein}
One of the ``Eleven Science Questions for a New Century'' listed in the U.S. National Academies' report ``Connecting Quarks to the Cosmos'' [1] was``Did Einstein have the last word on gravity?''  Part of the proposed LISA mission's funding will come from the``Beyond Einstein'' project that NASA established to answer this question.  
One of the four major goals of the LISA mission is to detect and measure extreme-mass-ratio inspirals (EMRIs), the inspirals of neutron stars or solar-mass blackholes (mass around 10 solar masses) into super-massive black holes (mass greater than $10^6$ solar masses).  Accurately measuring an EMRI is valuable for at least two reasons:  First, it ``maps out'' the spacetime around a black hole.  Second, it also makes it possible to perform precision tests of general relativity, including Ryan's quadrupole-moment test of the black-hole uniqueness theorems [2].  In this paper, we focus particularly on testing general relativity by performing the Ryan test in a theoretically robust and reliable manner. 

\subsection{Using IMBHs to perform the Ryan test in a theoretically robust manner}
Recently, Miller [3] has pointed out that if IMBHs exist, their inspirals, as well as those of EMRIs, can be used to perform the Ryan test.  The inspirals of IMBHs into larger IMBHs or into super-massive black holes (SMBHs) generate much stronger gravitational waves than do EMRIs and are detectable at greater distances.  These intermediate-mass-ratio inspirals (IMRIs) are also observable  earlier in the inspiral than are  EMRIs.  It is well known that the theoretical predictions based on perturbative [4] and post-Newtonian [5] techniques are based on assumptions and approximations that are fully justified only at these earlier stages of the inspiral [6-9]. Therefore, these predictions cannot be relied on during the later stages of the inspiral near the last stable orbit (LSO), or the innermost stable circular orbit (ISCO) in the circular orbit case. Unlike EMRIs,  IMRIs can be observed in the theoretically secure earlier stage of the inspiral, in addition to the theoretically uncertain later part near the ISCO or LSO. We study the feasibility of using only this earlier, more theoretically well-predicted, part of the inspiral to perform the Ryan [2] test.

\subsection{Detecting usable IMRI signals}
Some IMRI signals are far stronger than any EMRI signals, and extracting them therefore requires far less computing. 
EMRI signals can only be discovered by a matched-filter search that requires a substantial integration time and a huge template count, and may also require a two-stage hierarchical search.  In contrast, many IMRI signals can be discovered with a single-stage matched-filter search that covers only a short part of the inspiral or even just uses Fourier transforms instead of more complex Wiener filters, as Miller has demonstrated [3].  Not all IMRIs will be this strong, however, and there are some advantages, principally greater precision, to signals from EMRIs or from IMRIs with larger mass ratios. Therefore we carefully consider how to detect low-mass IMRIs, and compare their detectability to that of EMRIs with equal chirp masses.  We verify that some IMRIs are at least as detectable as previously considered EMRIs, and also provide sufficient data from the theoretically robust stage of the inspiral to perform a reasonably precise Ryan test.

\subsection{Verifying feasibility of proposed strategy}
We verify that it is possible to perform the Ryan test using only data from the early, theoretically robust part of the inspiral  by demonstrating that many IMRI chirp signals in this stage of their inspirals are: 
\begin{itemize}
\item 	in the LISA frequency band, near the maximum sensitivity band
\item 	expected to cross through this frequency band (.003-.01 Hz) in the last year of inspiral
\item 	observable in three years or less (although the longer the observation period, the better)
\item 	reasonably likely to occur, if IMBHs are abundant,
\item 	in the region of parameter space where theoretically robust predictions are available
\item 	detectable if EMRIs are detectable, and
\item 	capable of being used to perform the Ryan test  at a precision level of ten percent or better.
\end{itemize}
In summary: The inspirals of intermediate-mass black holes (IMBHs) provide a less precise, but more theoretically robust alternate approach to performing the Ryan test. 

\section{Comparison with previous work}

\subsection{Other work on IMBH inspirals}
Several other authors---for example Miller [3,10] and Will [11]--- have recently considered IMBH inspirals, focusing primarily on the stage near the ISCO.  Our work is consistent with the work of Miller, but by focusing on a different and theoretically more secure stage of the inspiral, we find a different mass combination  to be of greater interest.  While a $10^3$ solar mass ($M\odot$) IMBH inspiral into a $10^6$ $M\odot$ SMBH is near ideal for investigating the ISCO, as Miller showed, a $10^{2.5}$ $M\odot$IMBH inspiral into a $10^5$ $M\odot$ SMBH is better for performing a theoretically robust Ryan test.

\subsection{Other works on inspiral signal strength}
Our computations of the strength of IMRI signals are based in part on Will's equation 10 and are consistent with his figure 2.  They are also consistent with, and partly based on, works by  Finn and Thorne [12] and Barack and Cutler [13].

\subsection{Other work on theoretical robustness}
We are not aware of other work that points out that using IMRIs to perform the Ryan test allows us to use the theoretically more robust region in which the assumptions of the post-Newtonian and perturbation theory approaches can be fully justified.  The lack of robustness of the inspiral predictions in the part of the inspiral near the last stable orbit, however, has been known for some time.  See works by Poisson and Flanagan and Hughes [6-9].

\subsection{Other work on detectability and search techniques}					
All the works cited above under signal strength are also relevant here.  We also relied very heavily on the recent paper by Gair, et al. [14]. In this paper, we deal with detectability by assuming the good first order approximation that systems with equal chirp mass are equally detectable. We find that if EMRIs can be detected and IMRIs are abundant, then it is possible to detect IMRIs with equal chirp masses that allow a substantially restrictive and theoretically robust Ryan test to be performed.

\section{Existence and Abundance of IMBHs}

\subsection{New evidence for intermediate-mass black holes}	
Astrophysicists have recently discovered evidence, mostly in x-rays but also in some dynamical studies, that suggests that, in addition to stellar-mass black holes  and super-massive black holes, intermediate-mass black holes, (IMBHs) may also exist in considerable numbers [15,16].   Previous evidence had suggested that super-massive black holes in the centers of large galaxies had masses from $10^6$ to $10^9$ solar masses, and stellar-mass black holes had masses from 3 to 30 solar masses, but no significant evidence had been found for black holes at any intermediate mass [17]. The most persuasive evidence to date for large numbers of IMBHs consists of unexpectedly bright x-ray sources called ULXs, for ultraluminous x-ray sources.

\subsection{Definition of IMBHs}
The traditional definition of stellar-mass black holes cuts off around 30 or 50 solar masses. 
Super-massive black holes are thought to reside only in the centers of bulge-containing galaxies and have masses typically greater than $10^6$ solar masses.  For simplicity, however, we will refer to black holes with mass below 100 solar masses as stellar-mass black holes, those with mass from $10^2$ to $10^6$ solar masses as IMBHs, and those with mass above $10^6$ solar masses as super-massive black holes.	

\subsection{The hypothesis of IMBH abundance}
It was once believed that super-massive black holes (SMBHs) were present only in quasars, but now they are thought to be present in every bulge-containing galaxy.  Similarly, it is now starting to be suspected that IMBHs are much more common than previously thought.  If a significant percentage of the ULXs turn out to be IMBHs and the ratio of x-ray-quiet IMBHs to ULXs is similar to the ratio of bulge-containing galaxies to quasars, then there may be hundreds of IMBHs for each SMBH.  A similar ratio is deduced if all or a significant fraction of globular clusters contain an IMBH [15], or if IMBHs are common in dwarf galaxies. 

None of these possibilities has yet been firmly established, but we will assume for the balance of this paper that IMBHs will turn out to be at least an order of magnitude more numerous than SMBHs.  Finally, we note that if this hypothesis is true, there is a continuous spectrum of black-hole masses from stellar-mass black holes on up through IMBHs to SMBHs.  

\subsection{Possible dominance of IMRIs}
 Unlike EMRIs, which can only be detected out to about 1 gigaparsec, or perhaps even less if the integration time must be reduced because of the huge template count requirement mentioned above, most IMRIs are strong enough to be detected throughout the entire universe, i.e at a range of more than 4 gigaparsecs [17,18].  Furthermore, we will be able to see IMRIs well before the final coalescence, while EMRIs will only be detectable near the final merger.  Because of the greater signal strength of IMRIs and the resulting greater volume in which IMRIs can be observed, LISA may possibly see more IMRIs than EMRIs, if IMBHs exist in the numbers suggested by the hypotheses just discussed, despite the fact that there are far more stellar-mass black holes than IMBHs [17].

\section{The frequency evolution function and the Ryan test}

\subsection{The frequency evolution function}
The basic observable gravitational wave form is quasi-sinusoidal with a slowly rising frequency, called a chirp.  By matched filtering, we can determine the frequency of the chirp as a function of time with an error of less than one tenth of a cycle.  This frequency evolution function (FEF) will be observed with this accuracy over hundreds of thousands of cycles---or even a million or more---in a typical chirp observed by LISA.  Hence the FEF will be known with an accuracy better than one part in $10^5$ or $10^6$.  This is what allows us to perform precision tests of general relativity, by comparing the observed FEF to a predicted FEF.

Many predictions of the FEF have been made, and they are most frequently expressed in terms of Taylor series expansions in a velocity variable v, which is closely related to the cube root of the observed frequency, or similar variables such as the eighth root of the time remaining in the inspiral. See recent reviews by Blanchet[5] and Sasaki and Tagoshi [4] for detailed expositions. We use v throughout the remainder of this paper, and hereafter, we assume basic familiarity with this well-known variable and these series expansions of the FEF and closely related functions.   

\subsection{ Definition of the Ryan test of the black-hole uniqueness theorems}
According to the black-hole uniqueness theorems, in general relativity the only astrophysically possible neutral black hole is a Kerr black hole, which is uniquely determined by its mass, M, and spin, S.  General relativity predicts that the magnitude of the suitably defined quadrupole moment, Q, of a Kerr black hole is ${S^2/M}$.   If Q is not equal to ${S^2/M}$, general relativity is falsified.   

Ryan showed that one can determine the mass, the spin and all other multipole moments of a black hole from the FEF of an  inspiral chirp, and that one can determine the mass, M, the spin, S, and the quadrupole moment, Q, from just the first four terms in the Taylor expansion of the FEF in the extreme-mass-ratio circular-orbit case.  This is also true in the general case with arbitrary mass-ratio and non-zero eccentricity, if you can determine all the other astrophysical parameters involved.  Using this decomposition of the FEF to check whether or not ${S^2/M}$ is the test of the black-hole uniqueness theorems by Ryan's method, or the Ryan test. This is one of the easiest and cleanest tests for the correctness of general relativity and one of the most restrictive on possible alternate theories of gravity.  In principle, one needs only three numbers (M, S, Q) for this test.  

\subsection{Finding the FEF experimentally}
The theoretically predicted FEF has many terms, each with a complicated coefficient.  To perform the Ryan test, we must understand at least the first four terms, as well as  up to 17 astrophysical parameters that affect the FEF at least slightly. (These 17 parameters will be discussed in more detail in a paper under preparation.  They are also discussed in Gair, et. al. [14]).  For observational purposes, however, we can merely regard the FEF as a Taylor series in v (or $(t-t_0)^{1/8}$ where $t_0$ is also an unknown) with unknown coefficients.  We just systematically guess the coefficients (including $t_0$) and use the Wiener matched filtering technique to test whether or not we guessed correctly. 

\subsection{The FEF Taylor series expansion}
In a convergent Taylor series, only a finite number of terms contribute significantly, and the other terms can be set to zero and ignored.  This is relevant to the theoretical robustness of the Ryan test calculation because only a finite number of terms have yet been calculated.  A few terms are known up to order 11 in v.  Most terms are known up to order 7, but some complex combinations of terms---particularly terms involving high eccentricity or self force---are still unknown at orders less than seven. You can safely discard or ignore higher-order terms if you can provide evidence that terms beyond a given order are probably small, although you can never prove that an unknown term is small.  The magnitude of the known terms in the FEF expansion (or a closely related expansion, the energy radiation function {dE/df}) seems to be reasonably well approximated by a simple exponential in n, i.e.  $(2 v)^n$.  If we assume that this approximation holds for the unknown terms, we can infer when it is safe to ignore these unknown terms.  This approximation suggests we can ignore terms beyond $n=7$  (seventh order) when $v<.18$.  Similarly it suggests ignoring terms beyond 6th order if $v <.15$, terms beyond fourth order if $v <.05$,  and terms beyond third order if $v <.02$.

\section{Robustness and Accuracy of the Ryan test}

 \subsection{The problem of robustness} 
It is also well known that the perturbative and post-Newtonian series-expansion-based predictions  of the FEF become less and less reliable as v gets larger and the binary gets closer to its final plunge and merger. (The plunge happens at $v=.4$ in the Schwarzschild case, $v=1.0$ in the extreme corotating Kerr case, but only $v=.33$ in the extreme counter-rotating Kerr case.) This decreased reliability is in part due to the lessening validity of approximations such as adiabatic inspiral and in part  due  simply to the slower convergence of the series expansions as v gets larger.  This problem is sometimes codified by renaming the late inspiral as the merger phase, and stating that we can predict the inspiral but not the merger.  Flanagan and Hughes [8] select a value equivalent to $v=.27$ to mark this boundary.  It is clear that somewhere between $v=.1$ and $v=.3$, the series expansions can no longer be relied on.  Based on comparisons of truncated higher-order expansions performed by Poisson and the graphs he prepared [6], it appears that accuracy up to the fourth order (the order needed for the Ryan test) can be guaranteed up to $v=.15$ or $v =.16$.  Based on the graphs in Poisson, we select the value $v =.16$ as a very conservative limit of reliability, and use $v<.16$ as the requirement for robustness and reliability for the balance of this paper.  

\subsection{Superior robustness of IMRIs}
This allows us to explain in detail the advantage of IMRIs over EMRIs in increased robustness.  The data for performing the Ryan test for a typical EMRI all comes from the part of the chirp where $v>.2$, a region that is suspect for the reasons briefly discussed in the preceding subsection.  
In contrast, it is actually possible to perform the Ryan test on IMRIs using only data from the part of the inspiral where $v<.16$, a part of the cycle that is much less subject to theoretical uncertainties.

\subsection{The potential accuracy of the Ryan test}
Reliably determining the accuracy of the Ryan test without a detailed study is not easy.  However, a reasonable estimate can be made rather simply.  As indicated above, the Ryan test is based on three numbers, M, S and Q.  M and S will be very well determined; the main uncertainty in the Ryan test will come from the uncertainty in the value of Q, which is primarily determined by its contribution to the fourth-order term in v, of which it constitutes approximately 50 percent for large spin cases.  

We can get a good estimate of the accuracy of the Ryan test by counting the number of cycles the fourth-order term contributes to the chirp.  Since we can measure cycles down to one tenth of a cycle or less, a contribution of 10 cycles may yield an ultimate accuracy in Q closer to one part in a hundred than to one part in ten. If we want a robust result, we consider only the part of the chirp with $v < .16$.  This entirely eliminates EMRIs.  (EMRIs typically have many cycles---often hundreds or thousands---contributed by the fourth-order term, but they all come from the part of the inspiral with $v>.16$) In contrast, most IMRIs have at least ten cycles between $v=.12$ and $v=.16$.  In many IMRI cases, the fourth-order term contributes over 100 cycles in this range.  This is illustrated in figure 1, where in all the systems falling below the thick dashed line, the fourth-order term contributes more than 10 cycles, while in those falling below the medium dashed line, the fourth-order term contributes over 100 cycles from $v=.12$ to $v=.16$, and in those falling below the thin dashed line, the fourth-order term contributes 1000 cycles from  $v=.12$ to $v=.16$. 

\subsection{Chirp-mass equivalence}  
It is well known that inspirals from systems with the same chirp mass, where chirp mass $= M^{2/5}m^{3/5}$ , are indistinguishable to the first order [19].  We say systems with the same chirp mass are chirp-mass equivalent.  At any given frequency, chirp-mass equivalent systems should be equally detectable to the first order . 
In Figure 1 below, the thin solid lines are lines of the same chirp mass.  From the top down, they are chirp-mass equivalent to an EMRI system consisting of a $10^6$ solar mass $M\odot$  black hole and, respectively
\begin{itemize}
\item 	a 100 $M\odot$ black hole
\item 	a 30 $M\odot$ black hole
\item 	a 7.0 $M\odot$ black hole
\item 	a 1.4 $M\odot$ neutron star
\item 	a .6 $M\odot$ white dwarf
\end{itemize}

The thick solid line at the top is the equal mass limit.  There are no systems above this limit.   The thick dashed line is the line of at least 10 cycles contributed by the fourth-order term in the FEF; systems above this line are not useful for a precision Ryan test.  The medium dashed line is the line of 100 fourth-order cycles contributed.  The thin dashed line at the bottom is the line of 1000 fourth-order cycles contributed.  The thin vertical line at $lg(M)=5$ is an approximate marker of the robustness boundary.  Systems with $lg(M)=4.7$ have $v<.16$ when they exit the LISA best- sensitivity band. Systems with $lg(M)=5.2$ have $v <.16$ when they enter the LISA best-sensitivity band.  Systems on the left side of this boundary are useful for performing tests, including the Ryan test, in the theoretically robust zone.

\section{Conclusions}

\subsection{Conclusions with regard to possibility of a theoretically robust Ryan test.}  
Substantial numbers of detectable IMRI systems should be capable of being used to perform the Ryan test, assuming that IMBHs exist and are somewhat more numerous than SMBHs.  A calculation at roughly the ten percent level should be possible with one year's data on an appropriate system, and a calculation ten times more precise should be possible with five or more years of data on a favorable system. A glance at Figure 1 establishes that all EMRIs in the LISA band are outside of the theoretically robust region, and that many IMRIs are inside the theoretically robust region.
 
\subsection{Conclusions with regard to practicality}
This is a completely doable strategy for performing the Ryan test in a theoretically robust manner, implementable with theoretical calculations that have mostly already been done.  It can be done with all IMRIs whose larger mass M is less than ${10^5}$ solar masses, and whose smaller mass m is  100 or more times less than M.  These systems will almost surely be detectable if m is greater than some boundary mass between 100 and 1000 solar masses, whose exact value is yet to be determined.  These IMRIs will provide data accurate enough to do a Ryan test to at least $10\%$ precision.

\subsection{ Summary}
IMRIs may exist in large numbers.
If they exist, they will be very visible to LISA and at least as detectable as EMRIs.
They will allow us to perform the Ryan test to at least 10 percent accuracy in the analytically secure regime $v <.16$.

\begin{figure}[htbp]
\includegraphics[totalheight=9cm]{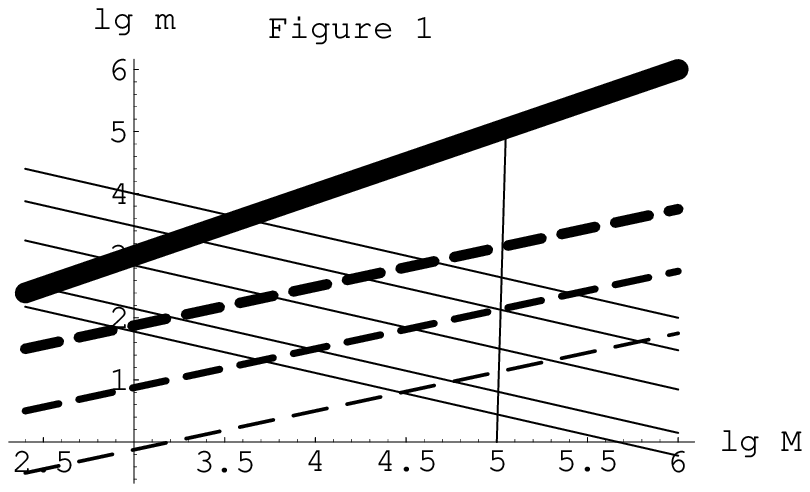}
\end{figure}

\end{document}